\documentstyle[11pt,newpasp,twoside,epsf]{article}
\begin{document}

\title{Neutron Stars in Supernova Remnants}

\author{Franco Pacini}
\affil{Arcetri Astrophysical Observatory and\\
Department of Astronomy and Space Sciences\\
University of Florence,
Largo E. Fermi 5,
I-50125, Firenze, Italy}

\begin{abstract}

In the following, we shall briefly summarize some facts and ideas
concerning the
presence of neutron stars in Supernova remnants. 
While sources similar to the Crab Nebula require the presence of a
central energetic object, shell-type remnants such as Cas A are
compatible with the presence of neutron stars releasing a weak
relativistic wind.
\end{abstract}
\parskip=0pt
Supernova remnants are usually classified into two extreme categories:
shell-type
and filled-center (plerions).  In the case of shell remnants, the
edges of the 
source appear bright, the interior rather faint.  The typical radio
spectrum is 
steep ($S_\nu \propto \nu^\alpha$ with $\alpha \simeq - 0.5$) and is due to synchrotron
radiation from relativistic electrons produced by shocks in the region where the expanding
debris interact with the circumstellar/interstellar medium.  Cas A is the
prototype of shell-type remnants.

On the opposite end, the Crab Nebula has been assumed to be the
typical plerion,
where a central neutron star continuously converts its rotational
energy into a
magnetized relativistic wind.  This wind expands and produces a
center-filled 
nebular emission which has a rather flat radio spectrum 
($\alpha \simeq - 0.2$).  It is not 
surprising that several remnants show both characteristics (internal
emission and 
bright limbs) since some plerions expand into a relatively dense
medium (composite
remnants).  For more details we refer the reader to a broad review by
Frail (1998) and to a full coverage of the subject in the
Proceedings of the Arcetri - Elba Workshop ``Relationship between
Neutron Stars and Supernova Remnants'' (Bandiera et al. 1998).
We also refer to two accompanying papers in these
Proceedings which deal with the Crab Nebula (Amato 1999) and with
plerions in general
(Bandiera 1999).  
Among the 215 catalogued Supernova remnants (Green 1996) 85\%
are of the shell-type.

In the past, it was widely held that plerions contain a neutron
star while shell-type remnants do not, either because the explosion blows apart the
entire star 
or because the central object becomes a black hole.  The basis for
this belief was the lack of evidence for an internal radio pulsar and/or of a
source of 
relativistic wind.  On the other hand, already in the early work on pulsars,
it had 
been suggested that magnetic fields of neutron stars can reach values up
to, say,
$10^{14} - 10^{15}$ gauss (Woltjer 1968).  The initial  loss of rotational
energy would
then be very fast and the central neutron star would soon become
unable to produce a strong relativistic wind. 

In this framework the possible presence of a neutron star hidden in
Cas A (with 
period $P \geq 0.7$ sec $ B > 10^{14}$ gauss) was discussed by Cavaliere and
Pacini (1970).

It is also worth recalling that the estimate for the rate of core
collapse Supernovae (around one every 30-50 years) is about a factor two larger than
the same 
estimate for the birthrate of radiopulsars (about one every 100
years). Despite the uncertainty of these estimates,  
this suggests the existence of a large fraction of neutron stars which do not
appear as radiopulsars (Helfand and Becker 1984, Helfand 1998).

Indeed, in recent years the observational evidence for the presence of
neutron stars in Supernova remnants has changed, largely because of
observations in the X and $\gamma$-rays bands.
According to Frail (1998), at least 19 Supernova
remnants contain neutron stars which manifest themselves in different ways.
Seven of them 
are classical radiopulsars (only one third of the total!);  3 are
X-ray binaries;  2 are slow X-ray pulsars;  2 are soft gamma-ray 
repeaters;  5 are
radio quiet 
neutron stars (detected because of their thermal X-ray emission).
Although this list is not updated, the picture has not changed substantially in
its implications.

The reasons which lead to different manifestations of neutron stars
in Supernova
remnants are not fully understood.  However, the binary nature of
the system (i.e. the role of accretion) or the initial 
period of the neutron star
and/or the 
magnetic field strength are likely to play a dominant role in
determining the evolution.  These factors may not be 
independent.  In particular the
initial rotation of neutron stars could be related to the strength of the
coupling between the collapsing core and the stellar envelope 
(Tsuruta and Cameron 1966,
Pacini 1983).  Strong magnetic fields would lead to slow initial 
rotation.  Even
if this 
coupling is not important, the wind produced by a newly born, strongly
magnetized, neutron star could damp the rotation immediately
or shortly after the
explosion.  As a consequence, some young remnants could contain a slow neutron
star, unable to support a strong wind.  The evidence (from the slowing
down
rate) for the presence of neutron stars with periods of several seconds
and fields up to and above $10^{14}$ gauss 
in some remnants suggests
that this occurs rather
frequently.  This could lead to the dominance of the stellar magnetic energy 
$ \simeq B^2 R^3$ over the rotational energy $ \simeq M\ R^2 \Omega ^2$,
a so-called ``magnetar''.  
Magnetic
fields would then become the main energy source, which could be released, e.g.
through flares.  This possibility, already envisaged by Woltjer (1968), has been
revived in 
more recent times as a possible explanation of the soft gamma-ray
repeaters (Duncan and Thompson 1992).  We stress however that the standard
determination of the field value is based upon the energy
loss in a dipole field:
any deviation from the dipolar geometry inside the speed of light cylinder (or,
even worse, the existence of alternative slowing down mechanisms) would
invalidate the determination of the strength on the stellar surface.

In any case it is now evident that the lack of a radiopulsar or of the
plerionic
component cannot be taken as proof for the non existence of a neutron
star inside Supernova remnants.   In this case only observations can tell us
whether the 
central object is a slow rotator with a strong magnetic field or, at
the opposite end, a fast but weakly magnetized neutron star (or any other
combination of rotation and field strength unable to provide a strong
pulsar wind).

Finally, we note that SN 1987A does not show, in its light curve,
evidence for a
pulsar energy input larger than  $2 \times 10^{36}$ ergs ${\rm s}^{-1}$  (Danziger, private
communication).  This lends additional support to the idea that some remnants
may contain neutron stars less energetic than those usually present in plerions.

\vskip 1cm
{\bf Note added} (October 1999).
After I.A.U. Symposium 1995 took place, a central point source
(neutron star?) in
Cas A has indeed been found by the AXAF - Chandra X-ray satellite,
shortly after its launch
(Tananbaum et al., 1999).  Observations in the various bands are currently
underway.  These will be able to provide more information about the nature and
possible periodicity of this source, as well as about the 
correctness of the points
discussed above.

\vskip 1cm
\acknowledgements

  I am indebted to E. Amato, R. Bandiera, M. Salvati, L. Woltjer for many
  discussions on the subject of this talk.  This work was partly supported
  by the Italian Space Agency.

\end{document}